\documentclass[aps,twocolumn,showpacs,preprintnumbers,amsmath,amssymb,amsfonts]{revtex4}
\usepackage{amsmath,amssymb,amsfonts}
\usepackage{dcolumn}% Align table columns on decimal point
\usepackage{bm}% bold math
\usepackage{verbatim}
\usepackage{epsfig,graphicx}
\usepackage[english]{babel}

\makeatletter
\newcommand{\const}{\operatorname{const}}
\newcommand\gsl{\ifmmode\textsl{g}\else g\fi}

\makeatother

\begin{document}

\preprint{APS/123-QED}

\title{Influence of cooling on dynamics of buoyant jet}
\author{V.~P.~Goncharov$^1$}
\author{V.~I.~Pavlov$^2$}

%%\ead[url]{home page}
%% \fntext[label2]{}
%% \cortext[cor1]{}
%%\address[label1]{A. M. Obukhov Institute of Atmospheric Physics PAS, 109017 Moscow, %%Russia\fnref{label3}}
%%\fntext[label3]{$^{1}$}

\affiliation{$^{1}$A. M. Obukhov Institute of Atmospheric Physics PAS, 109017 Moscow, Russia}

\affiliation{$^{2}$Univ.~Lille, UFR des Math\'ematiques Pures et Appliqu\'ees, CNRS~FRE~3723~-~LML, F-59000
Lille, France}

%UFR des Math\'ematiques Pures et Appliqu\'ees -- LML CNRS UMR 8107, \\
%Universit\'e de Lille 1, 59655 Villeneuve d'Ascq, France}

\date{Received}

\begin{abstract}
The Rayleigh--Taylor instability which is responsible for the occurrence of narrow upward jets are studied in the scope of the nonhydrostatic model with horizontally--nonuniform density and the Newtonian cooling. As analysis shows, the total hierarchy of instabilities in this model consists of three regimes -- collapse, algebraic instability, and inertial motion. Realization of these stages, mutual transitions and interference depend on a ratio between two characteristic time scales -- collapse time and cooling time.
\end{abstract}


\maketitle

\section{Introduction}

It is known that the Rayleigh–Taylor instability (RTI) appears when a higher density fluid is positioned above a fluid with lower density in a gravitational field or in an accelerated system when the fluid with lower density accelerates the fluid of higher density. The physical cause of  the initial layered density stratification can be various. Layers with temperature inhomogeneity, layers of salinity, inhomogeneous distribution of bubbles can be observed in geophysical conditions. Turbidity currents, whose the average density stratification derives from suspended mud or silt, can also exist.

\begin{figure}[ht]
\center
\includegraphics[width=0.45\textwidth]{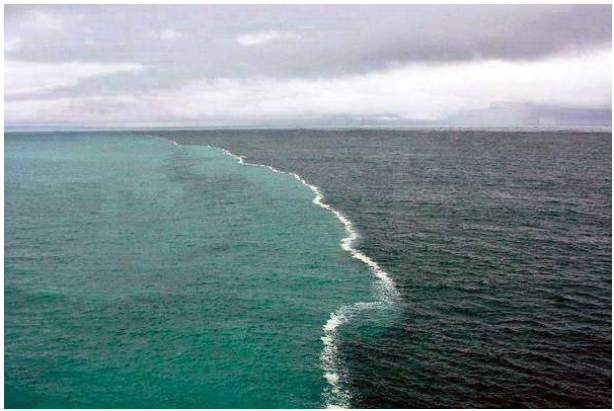}
\caption{A meeting of two oceans in the Bering Strait region \cite{bs14} (the relative velocity of flows is zero). A sharp boundary between regions with different temperatures is clearly observed. This can be an example of the beginning development of the RTI of the interface of regions with different horizontally distributed relative "buoyancy" $ \tau (\mathbf {x}, t) = a_{h} (\partial \varrho / \partial T) \Delta T / \varrho_{0} $. Here, $T$ is the temperature, $\varrho$ is the density of water, the parameter $a_{h} $ is the horizontal component of non-inertial (for example, centrifugal) acceleration.
} \label{TheMeetingOfTwoOceansInTheGulfOfAlaska}
\end{figure}

\begin{figure}[ht]
\center
\includegraphics[width=0.45\textwidth]{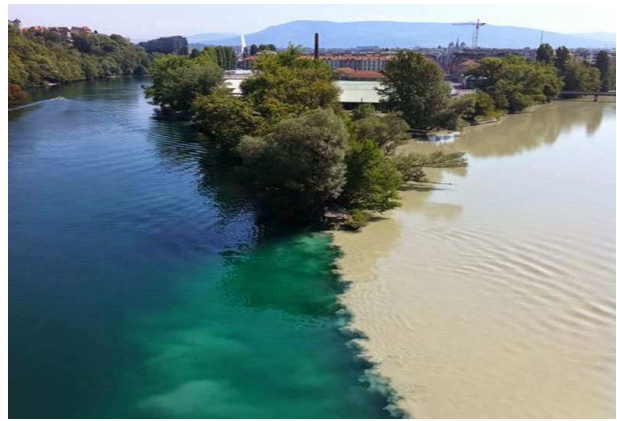}
\caption{A meeting of flows of Rhone and Arve (France) \cite{ra14}: the relative velocity of flows is zero. A sharp boundary divides regions with different thermodynamical properties.  In the right zone in Figure, the field $\tau(\mathbf{x},t)$ depends on the concentration of suspended particles of mud or silt.} \label{Rhone-ArveMeeting}
\end{figure}

The main aim of this work is to study the influence of cooling on the development of the RTI. In astrophysics, geophysical fluid dynamics and technical applications, numerous examples can be found when RTI is initiated by thermal irregularities localized in thin horizontal (perpendicular to gravity) layers. In boundary layers, such irregularities arise as a result of nonuniform heating and look like islets of more hot (light) fluid.

The dynamics of these thermal islets is essentially nonlinear because their occurrence severely disturbs the balance in the fluid. (Let us remind that when a system is conservative and its initial state coincides or close to the unstable equilibrium then, at an early stage, its perturbations should evolve in the regime of exponential growth. Such behavior is predicted by the linear theory. That is how the instability develops in the classical problem~\cite{ray83} about the evolution of a plane interface separating two fluids when a heavier fluid lies above a lighter one in the gravity field. Traditional vision of the process in terms of spectral modes is that a single--mode (sinusoidal) initial perturbation (of small magnitude) of the unstable horizontal interface between fluids starts in the exponential regime (described by linear differential equations for interface deformation), proceeds to the nonlinear regime, and finally enters a turbulent regime where multiple space scales emerge.) For the thermal islets, at least at an early stage as long as vertical motions are so small that the hydrostatic balance is weakly broken, to our mind, a more plausible is another model. Such model can be constructed within the framework of the so--called "shallow water approximation" with horizontally nonuniform density~\cite{gp12,gp13a,gp13b}: in fact, by describing the process of RTI, one should distinguish between two limit cases based on the ratio of characteristic thickness $l$ of the fluid layers and the characteristic horizontal space scale of initial deformation of the interface $L$: "deep" $l / L \gg 1$ and "shallow" $l / L \ll 1$ fluids.

A quite full understanding of islets dynamics can be gained in the framework of a minimal model (see below) by analyzing its symmetries without any solving a Cauchy problem. As shown in Appendix, the existence of a scaling symmetry means that in addition to total energy $H$ the model conserves one more integral $G = \frac{1}{8}(dI/dt)^{2}-HI$, where $I$ is the moment of inertia of the islet proportional to the squire of the characteristic size of islet $\langle r \rangle $ multiplied on its vertical magnitude $\langle h \rangle$. The quantity $I$ serves as an indicator of the collapse when $\langle r \rangle \rightarrow 0$ and $\langle h \rangle \rightarrow \infty$, the integrals $H$ and $G$ determine the characteristic time of collapse $t_0$.

The existence of integrals $H$ and $G$ permits to understand the cause of appearing of the collapse of islets. In fact, differentiating $G$ with respect time $t$ at once gives us the so--called virial theorem $d^{2}I/dt^{2}=4H$. The sufficient criterion of the collapse, which immediately follows from the virial theorem, says that the positive quantity $I$ vanishes in a finite time for any initial states corresponding to a negative constant $H$. By virtue of incompressibility, this indicates that islets of more light fluid are collapsing into a vertical axis, forming infinitely narrow (singular) upward jets. If jets develop from quiescent states, their height grows as $h\sim(t_{0}-t)^{-1}$ and reaches infinity in finite time $t_{0}=\sqrt{G/(8H^{2})}$.

There is an extensive literature (see for example Ref.~\cite{zk11} and references therein) devoted to the phenomenon of collapse, also referred as the "blowup". This phenomenon is a sufficiently universal mechanism by which instabilities manifest themselves in nonlinear physical systems~\cite{ef09,gp08,gon09,gon11,gp06,gp07,gp10,gp12}. That is why it seems logical to assume that collapses can be key to the understanding of strong turbulence~\cite{kuz04,knnr07,gp14}.

It is obviously that scenario of self-similar collapse is an idealization. In fact, the closer is a system to its collapse, the more is its deviation from hydrostatic approximation. For this reason, as shown in~\cite{gp15a,gp15b} (see also Fig.~\ref{fig3}, taken from~\cite{gp15b}), under the action of nonhydrostaticity the regime of blow-up instability slows down and goes into the regime of algebraic instability.

The salient features of algebraic instability can be understood by using dimensional analysis. Suppose that along with the characteristic interval of possible singularity formation $t_0$, the problem has also other parameters. Among them are gravity acceleration $\gsl$, the jet height $h$, its initial value $h_{0}$, the current time $t$, and the Atwood�s number $A$. Thus, $h=h(t,\gsl,t_{0},h_{0},A)$, which means that only the three-argument functional dependence $h = h_0 F (\gsl t^2/h_0, t_0/t , A)$ is admitted for a dimensionless combination. At large time $t\gg t_{0}$, the second argument tends to zero. In this case, the system must "forget" its initial state and dependence on $h_0$ must disappear. This is possible only if the function $F$ is linear in the first argument. As a result, these qualitative arguments lead us to the law $h\sim\gsl t^{2}f(A)$.

The minimal nonhydrostatic model allows to describe not only the initial and algebraic stage (see asymptotics $a$ and $b$ in~Fig.~\ref{fig3}) but also the transition between them. Both these regimes are the results of the self-similar development of upward jets arising under influence RTI in neglecting by dissipative effects. However, the number of open questions becomes considerably larger if we consider more general models, which use other boundary and initial conditions and take into account effects of viscosity, diffusion, thermal and electric conduction. All these reasons lead to the occurrence of new typical scales of motion.

Without any doubt, the most important of all dissipative effects are those affecting the buoyancy force. In this context, thermal losses due to the cooling are more important in comparison with losses due to viscosity. Indeed, thermal smoothing due to cooling leads to density homogenization. For the warm upward jets, it means that eventually they become neutrally buoyant and hence move inertially -- with a constant velocity without the action of any external forces.

Thus, if the nonhydrostatic model is supplemented by the cooling effect, it should predict the inertial regime when the upward jet height grows as $h\sim h_{\ast}+c_{\ast}t$. The influence of model parameters on quantities $h_{\ast}$ and $c_{\ast}$ is one of the subject matters of this work.

This paper is organized as follows. In Sec. II we discuss the model setup and formulate the governing equations. In Sec. III we analyze self-similar solutions and consider possible scenarios of their behavior. In Sec. IV we summarize our results. Appendix A is devoted to scaling symmetry which is responsible for the self-similarity at the initial stage of development of RTI.

\section{Nonhydrostatic model with newtonian cooling}

\begin{figure}[t]
\center
\includegraphics[width=0.45\textwidth]{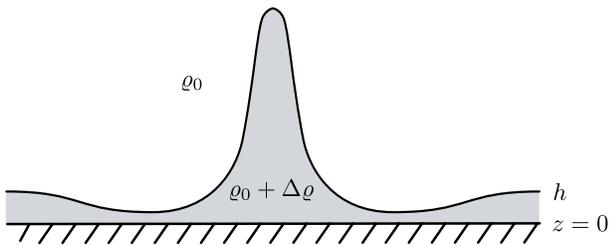}
\caption{The model of the active (lower) layer with horizontally-nonuniform density. This conservative model supposes that two incompressible fluids are separated by the interface $z=h(\mathbf{x},t)$ and subjected to the action of gravity $\gsl$. The upper fluid with the density $\varrho_{0} = \const$ expands to $z=\infty$, whereas the lower fluid with the density $\varrho_{0}+\Delta\varrho(\mathbf{x},t)$, where $\Delta\varrho/\varrho_{0}\ll1$, rests on the horizontal bottom $z=0$.} \label{fig2}
\end{figure}

We considered in previous works~\cite{gp15a,gp15b} an estimation two-layer (Fig.~\ref{fig2}), non--hydrostatic, model with depth--averaged flow in the active (lower) layer, which governed by the set of equations
%The nonhydrostatic model~\cite{gp15a,gp15b} of active layer is described by the equations
\begin{gather}
\partial_{t}\mathbf{u}+\left(\mathbf{u\cdot\nabla}\right)
\mathbf{u}=-\frac{1}{2h}\mathbf{\nabla}(h^{2}\tau)
-\frac{1}{3h}\mathbf{\nabla}\left(h^{2}\frac{d^{2}h}{dt^{2}}\right),\label{eq:1a}\\
\partial_{t}h+\mathbf{\nabla\cdot}\left(h\mathbf{u}\right)=0,\label{eq:1b}\\
\partial_{t}\tau+\mathbf{u\cdot\nabla}\tau=0. \label{eq:1c}
\end{gather}
The notations in Eqs.~(\ref{eq:1a})--(\ref{eq:1c}) are as follows: $\mathbf{x}$ are the Cartesian horizontal coordinates; $\mathbf{\nabla}$ is the horizontal gradient operator; $\partial_{t}$ and $d/dt$ are the partial and total time derivatives; $\mathbf{u}(\mathbf{x},t)$ is the depth averaged horizontal velocity in the active layer, $\tau(\mathbf{x},t)=\gsl\Delta\varrho/\varrho_{0}$ is the relative buoyancy, which, unlike $h$, may take any sign.

In the simplest case $\tau=\gsl=\const$, Eqs.~(\ref{eq:1a})--(\ref{eq:1c}) are reduced to the well-known Green--Naghdi equations which describe gravity waves on the surface of shallow water in the non-hydrostaticity approximation \cite{per67,gn76,wu81,ch92,chh94}. If the hydrostatic balance is broken sufficiently weakly, there is every reason to ignore the Green--Naghdi correction -- the last term in Eq.~(\ref{eq:1a}). It is in this approximation that Eqs.~(\ref{eq:1a})--(\ref{eq:1c}) were used to study the development of the Rayleigh-Taylor instability in large-scale flows with horizontally-nonuniform density \cite{gp12,gp13a,gp13b}. Nevertheless, the model (\ref{eq:1a})--(\ref{eq:1c}) is not a generalization of Ripa-type models \cite{rip93}. As shown in~\cite{gp14}, it is a parametrization composed of terms which dominate at the initial and next stage of the RTI.

We consider the new physical effect of the deceleration of the RTI process due to a possible buoyancy variation. Depending on the mechanism of buoyancy variations (cooling, phase transformations, and so on), there are various ways of its parametrization. For reasons of simplicity, we will use, instead of Eq.~(\ref{eq:1c}), the relaxation type equation
\begin{equation}
\partial_{t}\tau +\mathbf{u\cdot\nabla }\tau=-\tau/T_{c},\label{eq:2}
\end{equation}
where $T_{c}$ is the relaxation time scale for the Newtonian cooling. Note that the Newtonian cooling is an effective quantity which implies the total thermal losses including the thermal radiation. For real physical models, the radiation cooling can be used to evaluate the upper bound of relaxation time.

\section{Self-similar solutions (radial-symmetric case)}

In the radial symmetry case and the lack of rotation, Eqs.~(\ref{eq:1a}), (\ref{eq:1b}), (\ref{eq:2}) look as follows
\begin{gather}
h\frac{du}{dt}+\partial_{r}\left(\frac{1}{2}h^{2}\tau+
\frac{1}{3}h^{2}\frac{d^{2}h}{dt^{2}}\right)=0,\label{eq:2.1a}\\
r\frac{dh}{dt}+h\partial_{r}\left(ru\right)=0,\quad \frac{d\tau}{dt}=-\tau/T_{c},\label{eq:2.1b}
\end{gather}
where the radial component of velocity $u$, the height $h$, and the buoyancy $\tau$
depend on radial coordinate $r$ and time $t$ only.

It is more convenient to use Eqs.~(\ref{eq:2.1a}), (\ref{eq:2.1b}) in the Lagrangian representation. The direct way to do this is to consider the parametrization $r=\hat{r}(s,t)$, where $s$ is a new (lagrangian) coordinate such that
\begin{eqnarray}
u (r,t)|_{r=\hat{r}}=\hat{r}_{t} (s,t), \quad h (r,t)|_{r=\hat{r}}=\hat{h} (s,t),\nonumber\\
\tau (r,t)|_{r=\hat{r}}=\hat{\tau} (s,t).\label{eq:2.2}
\end{eqnarray}
Here and in the sequel the subscripts $t$ and $s$ will denote partial
derivatives $f_{t}=\partial_{t}f$, $f_{s}=\partial_{s}f$.

In these notations, Eqs.~(\ref{eq:2.1a}), (\ref{eq:2.1b})) rearranges to give
\begin{gather}
\hat{h}\hat{r}_{s}\hat{r}_{tt}+
\partial_{s}\left(\frac{1}{2}\hat{h}^{2}\hat{\tau} +\frac{1}{3}\hat{h}^{2}\hat{h}_{tt}\right)=0,\label{eq:2.3a}\\
\quad\partial_t\left(\hat{h}\hat{r}\hat{r}_{s}\right)=0,\quad
\hat{\tau}_{t}=-\tau/T_{c}.\label{eq:2.3b}
\end{gather}

Just like the nondissipative versions~\cite{gp12,gp13a,gp13b,gp15a,gp15b} of this model, Eqs.~(\ref{eq:2.3a}), (\ref{eq:2.3b}) possess compact self-similar solutions
\begin{gather}
\hat{h}=h_{0}\eta\sqrt{1-s^{2}},\quad\quad\hat{r}=\frac{h_{0}}{2\phi\sqrt{\eta}}s,\label{eq:2.4a}\\
\hat{\tau}=-\tau_{0}e^{-t/T}\sqrt{1-s^{2}}.\label{eq:2.4b}
\end{gather}
Here $\eta(t)$ is a time-dependent variable which must be found, $0\leq s\leq1$, and $h_{0}$, $\tau_{0}$ are positive constants fixed by initial conditions. These solutions describe a semi-ellipsoidal drop which rests on the horizontal bottom and, with time, turns into a thin jet.

If we impose an initial condition $\eta(0)=1$, then $h_{0}$ is an initial height of jet and $d_{0}=h_{0}/\phi$ is its initial bottom diameter. Thus, the quantity $\phi=h_{0}/d_{0}$ makes simple geometric sense of the aspect ratio at initial time.

Substituting of (\ref{eq:2.4a}), (\ref{eq:2.4b}) into (\ref{eq:2.3a}), (\ref{eq:2.3b}), we get
\begin{equation}
\frac{d}{dt}\left[\left(\frac{d\eta}{dt}\right)^{2}
\left(1+\frac{1}{8\phi^{2}\eta^{3}}\right)\right]-
3\frac{\tau_{0}}{h_{0}}\frac{d\eta}{dt}e^{-t/T_{c}}=0.\label{eq:2.5}
\end{equation}

Let the scale of time $T$ be chosen so that $h_{0}=\frac{3}{2}\tau_{0}T^{2}$. Then, after nondimensionalization, Eq.~(\ref{eq:2.5}) can be rewritten in the form
\begin{equation}
\frac{d\eta}{dt}\sqrt{1+\frac{1}{8\phi^{2}\eta^{3}}}=p,\quad
\frac{dp}{dt}\sqrt{1+\frac{1}{8\phi^{2}\eta^{3}}}=e^{-\alpha t},\label{eq:2.6}
\end{equation}
where $\alpha=T/T_{c}$.

Because our interest is only in solutions corresponding to quiescent states, Eqs.~(\ref{eq:2.6}) must be solved under initial conditions
\begin{equation}
\eta(0)=1,\quad p(0)=0,\label{eq:2.6a}
\end{equation}
while $\phi$ and $\alpha$ play a role of model parameters.

In order to find an upper estimation for the jet growth rate at the stage when cooling neutralizes buoyancy, it suffices to solve Eqs.~(\ref{eq:2.6}) approximately. Assuming that, as $t\rightarrow\infty$, $\eta\rightarrow\infty$, we get asymptotically
\begin{equation}
\eta\approx\eta_{\ast}+c_{\ast}t,\quad p\approx c_{\ast}.
\end{equation}
where $\eta_{\ast}$ and $c_{\ast}$ are integration constants dependent on both $\alpha$ and $\phi$.

\begin{figure}[t]
\begin{center}
\includegraphics[width=0.455\textwidth]{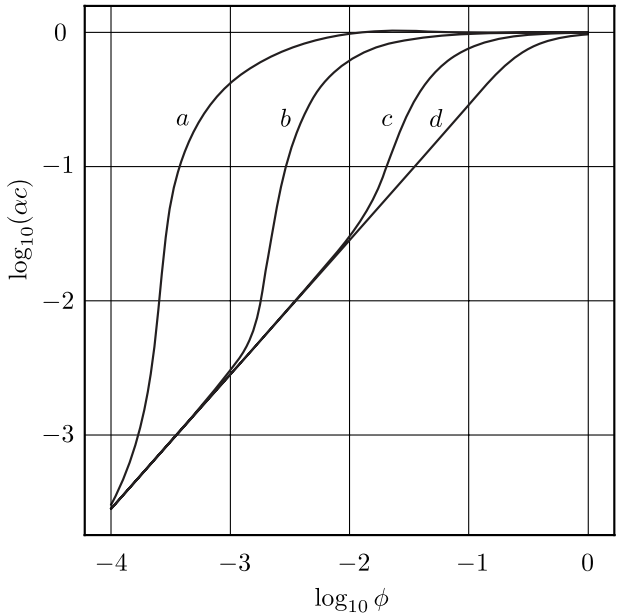}
\caption{The reduced growth rate $\alpha c$ of jet as the function of its initial aspect ratio $\phi$. The curves $a$, $b$, $c$, $d$ were calculated for $\alpha$ running through the values $0{.}001$, $0{.}01$, $0{.}1$, $1$.}
\label{fig5}
\end{center}
\end{figure}

If $\phi\gg1$, these constants can be found explicitly
\begin{equation*}
\eta_{\ast}=1-\alpha^{-2},\quad c_{\ast}=\alpha^{-1}.
\end{equation*}
Thus, as seen from Fig.~\ref{fig5}, in this case, the inequality $\alpha^{-1}\geq c_{\ast}$ majorizes the jet growth rate while the inequality $\phi/\alpha\leq c_{\ast}$ minorizes it.

All three stages of instability -- the collapsing regime, the regime of algebraic growth, and the inertial regime are recognized quite clearly if only the Newtonian cooling time $T_{c}$ much larger than the collapse time $t_{0}$. Since, according to Eq.~(\ref{eq:a13a})
\begin{equation*}
t_{0}=\sqrt{\frac{h_{0}}{6\tau_{0}\phi^{2}}}=\frac{T}{2\phi},
\end{equation*}
this condition implies that
\begin{equation}
2\phi\gg\alpha.\label{eq:2.9}
\end{equation}
If not, the algebraic regime is masked by inertial one.

\begin{figure}[t]
\begin{center}
\includegraphics[width=0.455\textwidth]{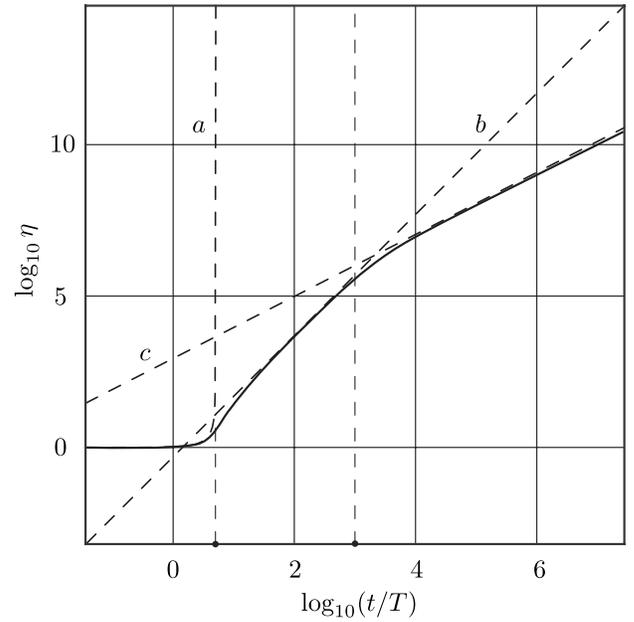}
\caption{Three stages of instability: the collapsing regime ($a$), the regime of algebraic growth ($b$), and the inertial regime ($c$). The regime of algebraic instability is contained between two vertical dashed lines. The calculation is performed for the parameters $\phi=0{.}1$, $\alpha=0.01$.}
\label{fig3}
\end{center}
\end{figure}

The case of a weak cooling when $T_{c}\gg t_{0}$ is shown in Fig.~\ref{fig3}. The collapsing regime develops at the initial stage and goes on as long as the inequality $\phi^{2}\eta^{3}<1$ holds true. In this approximation, (\ref{eq:2.6}), (\ref{eq:2.6a}) is converted into equations
\begin{gather}
\frac{d\eta}{dt}=2\sqrt{2}\phi p\eta^{3/2},\quad
\frac{dp}{dt}=2\sqrt{2}\phi\eta^{3/2},\label{eq:2.10}\\
\eta(0)=1,\quad p(0)=0,\label{eq:6.7b}
\end{gather}
which have the collapsing solution (see \cite{gp12,gp13a})
\begin{equation}
\eta=\left(1-4\phi^{2}t^{2}\right)^{-1}.\label{eq:2.11}
\end{equation}

This solution (see Fig.~\ref{fig3}) remains valid until $t$ approaches enough close to $t_{0}$. Next, in time interval $t_{0}<t<T_{c}$, the jet develops in the regime of algebraic growth. At this stage, as long as the cooling has a weak influence, the instability can be approximated by the equations
\begin{equation}
\eta_{t}=p,\quad p_{t}=1,\label{eq:2.12}
\end{equation}
which result in a power-law growth
\begin{equation}
\eta=t^{2}/2.\label{eq:2.13}
\end{equation}

\begin{figure}[t]
\begin{center}
\includegraphics[width=0.455\textwidth]{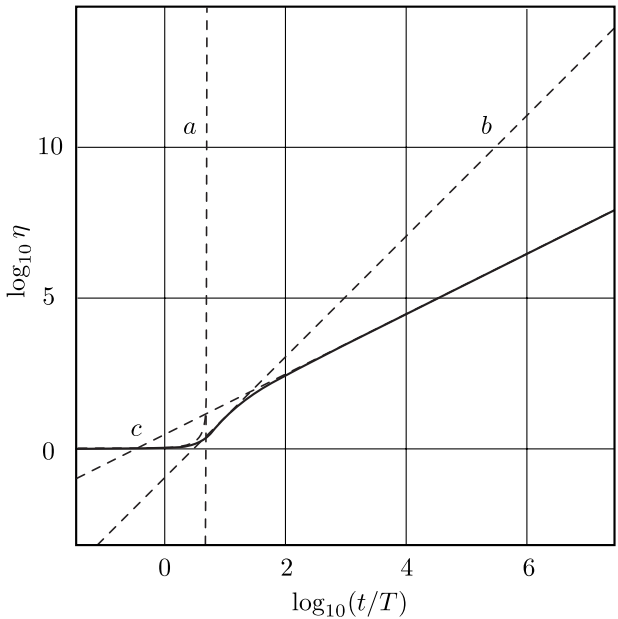}
\caption{The transition to the inertial regime without the stage of algebraic growth. The notations are the same as in Fig.~\ref{fig4} but the calculation is performed for the parameters $\phi=0{.}1$, $\alpha=0.1$.}
\label{fig4}
\end{center}
\end{figure}

If the cooling is so strong that times of $t_{0}$ and $T_{c}$ become rather close or even $T_{c}<t_{0}$, the transition to the inertial regime happens without the stage of algebraic growth. Just that very case is shown in Fig.~\ref{fig4}.

\section{Conclusions}

We now summarize the main results of the work. The main goal of this paper was to study the combined influence of non-hydrostaticity and of cooling process on the final stage of the RTI in the model of the active layer with horizontally-nonuniform density.

As shown in this paper, all three regimes (collapse, algebraic instability, and inertial motion) exist only for jets whose the initial aspect ratio $\phi$ is sufficiently large and the cooling is weak enough so that the inequality $1>2\phi\gg\alpha$ holds.

In this hierarchical sequence of regimes, the inertial motion is the only stage that is asymptotically achieved always in the presence of cooling independently of initial conditions. The realization of two preceding stages depends on initial conditions.

If initial conditions are such that the aspect ratio $\phi=h_{0} / d_{0}$ where $h_{0}$ is an initial height of jet and $d_{0}$ is its initial bottom diameter, becomes close to $1$, the collapsing regime has no enough time to develop. In this case, the existence and the lifetime of the second stage when the jet height undergoes a square growth in time depend on the relation between the collapse time $t_{0}$ and the cooling time $T_{c}$. The weaker is the inequality $T_{c}\gg t_{0}$, and hence $2\phi \gg \alpha$, the shorter is the second stage. As $T_{c}\lesssim t_{0}$, it is missing at all.

\begin{figure}[t]
\begin{center}
\includegraphics[width=0.455\textwidth]{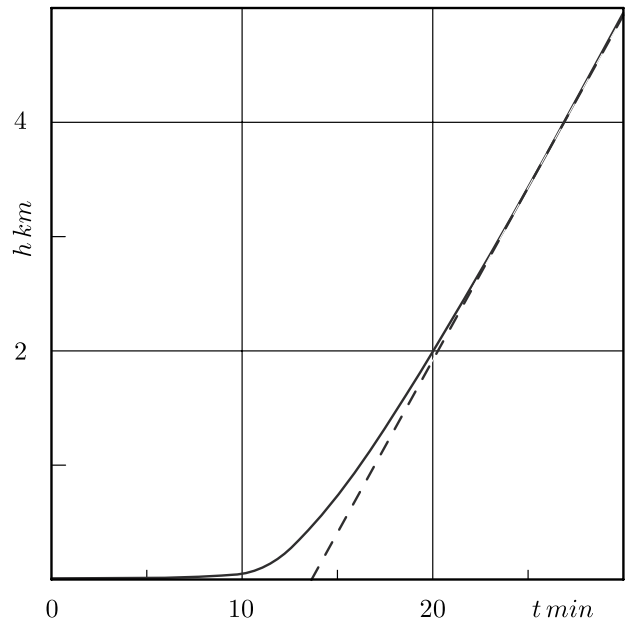}
\caption{Time-dependence of height of the jet under conditions typical of the Earth's atmosphere. The dashed line corresponds to the inertial regime.}
\label{fig6w}
\end{center}
\end{figure}

In order to assess the influence of cooling on the growth rate of jets, we consider conditions typical of the Earth's atmosphere. Assume that, in an initial state, a thermal islet has a semi-ellipsoidal shape Eq.~(\ref{eq:2.4a}) with the height $h_{0}=10\,m$ and the bottom diameter $d_{0}=1\,km$, as well as it is $\Delta T=2^{\circ}\mathrm{C}$ warmer than the ambient atmosphere. These values corresponds to the aspect ratio $\phi=0{.}01$ and the time scale $T\approx 10\,sec$. If the cooling time amounts $T_{c}\approx 5\,min$, then $\alpha = T / T_c = 0{.}03$, and our theory gives $c_{\ast}\approx 5\, m\cdot sec^{-1}$. As shown in Fig.~\ref{fig6w}, this growth rate is achieved at a height of $2\,km$ after approximately $20\,min$ when the thermal islet turns into a thin jet moving under inertia.
Note that, in given example the jet goes over into the stage of inertial motion by missing the intermediate stage of algebraic instability. This occurs because the cooling time $T_{c}\approx 5\,min$ is less than the collapse time $t_{0}\approx 8\,min$.

\section{Acknowledgments}

This work was supported in equal parts by the Russian Foundation for Basic Research (Project No.~15-05-00854), by the Presidium of the Russian Academy of Sciences (Program Fundamental Problems of Nonlinear Dynamics), and by the Russian Science Foundation (Project No.~14-27-00134).

\appendix
\section{Scaling symmetry of minimal model}

As a minimal model, we consider two-dimensional flow whose evolution is described by the equations
\begin{gather}
\partial_{t}\mathbf{u}+\left(\mathbf{u\cdot\nabla}\right)\mathbf{u}=
G_{0}\mathbf{x}-\frac{1}{2h}\mathbf{\nabla}\left(h^{2}\tau\right),\label{eq:a1}\\
\partial_{t}h+\mathbf{\nabla}\left(h\mathbf{u}\right)=0,\quad
\partial_{t}\tau+\left(\mathbf{u\cdot\nabla}\right)\tau=0. \label{eq:a2}
\end{gather}
These equations describe the depth-averaged flow in the active (lower) layer in hydrostatic SW approximation and can be formulated within the framework of the two-layer model that is shown in Fig.~\ref{fig2}.

This model supposes that two incompressible fluids are located on both sides of the interface $z=h(\mathbf{x},t)$ in the presence of gravity $\gsl$. The upper fluid with the density $\varrho_{0}$ expands to $z=\infty$, whereas the lower fluid with the density $\varrho_{0}+\Delta\varrho(\mathbf{x},t)$ is based on the horizontal bottom $z=0$.

The other notations are as follows: $\mathbf{x}$ are the Cartesian horizontal coordinates; $\mathbf{\nabla}$ is the horizontal gradient operator; $\partial_{t}$ and $d/dt$ are the partial and total time derivatives; $\mathbf{u}(\mathbf{x},t)$ is the depth averaged horizontal velocity in the active layer, $\tau(\mathbf{x},t)=\gsl\Delta\varrho/\varrho_{0}$ is the relative buoyancy, which, unlike $h$, may take any sign.
Note, depending on the sign of a constant $G_{0}$, the central force $G_{0}\mathbf{x}$ applied to the fluid can be both centripetal and also centrifugal. A more detailed derivation and discussion of this model is given in~\cite{gp12,gp13a,gp13b}.

Equations~(\ref{eq:a1}), (\ref{eq:a2}) are Hamiltonian and can be obtained from first principles with use of the Poisson brackets
\begin{gather*}
\{u_{i},u_{k}^\prime\}=h^{-1}\delta\left(\partial_{i}u_{k}-
\partial_{k}u_{i}\right),\label{eq:4}\\
\{h,u_{k}^\prime\}=-\partial_{k}\delta,\quad
\{\tau,u_{k}^\prime\}=-h^{-1}\delta\partial_{k}\tau,\label{eq:a4}
\end{gather*}
and  the Hamiltonian
\begin{equation*}
H=\frac{1}{2}\int d\mathbf{x}\left(h\mathbf{u}^{2}+ h^{2}\tau-G_{0}h\mathbf{x}^{2}\right).\label{eq:a5}
\end{equation*}
Here and in what follows, primed field variables mean the dependence on the primed spatial coordinates, $\delta=\delta(\mathbf{x-x^{\prime}})$ is the Dirac delta function, all the trivial Poisson brackets are omitted for the sake of space, and all the integrals are taken over the whole area occupied by the $2D$-flow.

In addition to $H$ the system has also other motion invariants. Some of them are annihilators of the Poisson brackets and are referred to as Casimirs. For the model (\ref{eq:a1}), (\ref{eq:a2}), total mass $Q$ and total buoyancy $N$
\begin{equation*}
Q=\int d\mathbf{x}\,h,\quad N=\int d\mathbf{x}\,h\tau,\label{eq:a6}
\end{equation*}
are the simplest of them. If a conserved functional $F$ is not the Casimir, then it is a generator of the symmetry transformation under which equations of motion are invariant~\cite{b87}.

In order to find such an invariant, we consider two integrals
\begin{equation*}
V =\int d\mathbf{x}\,h\,(\mathbf{x\cdot u}),\quad
I=\int d\mathbf{x}\,h\mathbf{x}^{2},\label{eq:a7}
\end{equation*}
which are treated as the virial and the moment of inertia.

As can be verified directly, being time-dependent quantities, integrals $V$ and $I$ obey the equations
\begin{equation*}
\frac{dI}{dt}=2V,\quad \frac{dV}{dt}=2(H+G_{0}I).\label{eq:a8}
\end{equation*}
These equations can be rewritten in the canonical form
\begin{eqnarray}
\frac{dI}{dt}=\frac{\partial G}{\partial V},\quad
\frac{dV}{dt}=-\frac{\partial G}{\partial I},\label{eq:a9}\nonumber\\
G=V^{2}-2IH-G_{0}I^{2}\label{eq:a10}.
\end{eqnarray}
where the functional $G$ plays the role of the Hamiltonian and hence is the new invariant of motion.

Once $G$ is a constant of motion, taking its derivative with respect to time gives
\begin{equation}
\frac{d^{2}I}{dt^{2}}=4\left(H+G_{0}I\right).\label{eq:a11}
\end{equation}
If $G_{0}=0$, from (\ref{eq:a11}) it follows that
\begin{equation}
\frac{d^{2}I}{dt^{2}}=4H.\label{eq:a12}
\end{equation}
In nonlinear optics~\cite{vpt71}, this relation is known as virial theorem. In this simple case, for solutions evolving from the quiescent states $\left(dI/dt\right)_{t=0}=0$, Eqs.~(\ref{eq:a10}), (\ref{eq:a12}) give
\begin{equation}
I=-\frac{G}{2H}+2Ht^2,\label{eq:a13}
\end{equation}
Thus, inequalities $H<0$ and $G\geq0$ are the necessary and sufficient condition for collapsing from quiescent states.

The collapse time $t_{0}$ is defined as time at which $I$ vanishes. If $G_{0}=0$, from this condition and Eq.~(\ref{eq:a13}) it is easy to find that
\begin{equation}
t_{0}=\sqrt{\frac{G}{4H^{2}}}=\sqrt{-\frac{I_{0}}{2H_{0}}},\quad
H_{0}=\frac{1}{2}\int d\mathbf{x}h^{2}\tau,\label{eq:a13a}
\end{equation}
where $I_{0}$ and $H_{0}$ are integrals calculated from the initial distributions $h$, $\tau$ on condition that $\mathbf{u}=0$.

In order to show that the invariant $G$ is responsible for the scale symmetry, we consider the transformation which changes both independent and dependent variables as follows
\begin{equation}
\mathbf{x}^{\prime}=\frac{\mathbf{x}}{\sqrt{I}},\quad
t^{\prime}=\int \frac{dt}{I},\quad
h=\frac{h^{\prime}}{I},\quad\tau=\tau^{\prime}.\label{eq:a14}
\end{equation}
Using the fact that
\begin{equation*}
\mathbf{u}=\frac{1}{\sqrt{I}}\left(\mathbf{u}^{\prime}+V\mathbf{x'}\right),\label{eq:a15}
\end{equation*}
it is easy to verify that transformation (\ref{eq:a14}) leaves Eqs.~(\ref{eq:a1}), (\ref{eq:a2}) invariant, namely, brings them to the form
\begin{eqnarray}
\partial_{t}^{\prime}\mathbf{u}^{\prime}+
(\mathbf{u^{\prime}\cdot\nabla^{\prime}})\mathbf{u}^{\prime}
=G\mathbf{x}^{\prime}-\frac{1}{2h^{\prime}}
\mathbf{\nabla}^{\prime}\left(h^{\prime2}\tau^{\prime}\right), \label{eq:a16}\\
\partial_{t}^{\prime}h^{\prime}+
\mathbf{\nabla}^{\prime}\left(h^{\prime}\mathbf{u}^{\prime}\right)=0,\quad
\partial_{t}^{\prime}\tau^{\prime}+
\left(\mathbf{u^{\prime}\cdot\nabla^{\prime}}\right)\tau^{\prime}=0, \label{eq:a17}
\end{eqnarray}
where $\mathbf{\nabla}^{\prime}=\partial/\partial\mathbf{x}^{\prime}$ and $\partial_{t}^{\prime}=\partial/\partial t^{\prime}$.
In doing so, the time-dependent integrals $I$, $V$  become invariants
\begin{equation*}
I\rightarrow I'=1,\quad V\rightarrow V'=0,
\end{equation*}
while the integrals of motion $H$, $G$ change as
\begin{equation*}
\quad H\rightarrow H'=-G,\quad G\rightarrow G'=G,
\end{equation*}
Obtained equalities should be considered as normalizing conditions in solving the problem (\ref{eq:a16}), (\ref{eq:a17}).

Note that in the special case of steady-state solutions when $\mathbf{u}^{\prime}=0$, all the conditions except for $I'=1$ are satisfied automatically and the problem (\ref{eq:a16}), (\ref{eq:a17}) reduces to the equation
\begin{equation*}
2Gh'\mathbf{x}^{\prime}=\mathbf{\nabla}^{\prime}\left(h^{\prime2}\tau^{\prime}\right).
\end{equation*}


\newpage
\begin{thebibliography}{99}


\bibitem[\protect\citeauthoryear{Two~oceans}{2014}]{bs14}
A meeting of two oceans in the Bering Strait region, 
\url{http://www.adn.com/article/mythbusting-place-where-two-oceans-meet-gulf\\-alaska}

\bibitem[\protect\citeauthoryear{Two~rivers}{2014}]{ra14}
A meeting of flows of Rhone and Arve (France), 
\url{http://whenonearth.net/meeting-two-rivers-rhone-arve\\-geneva-switzerland/}

\bibitem[\protect\citeauthoryear{Rayleigh}{1883}]{ray83}%1
Rayleigh, L.,
%Investigations of the character of the equilibrium of an incompressible heavy fluid of variable density,
Proc. Lond. Math. Soc. \textbf{14}, 170 (1883).
\url{http://dx.doi:10.1112/plms/s1-14.1.170}

\bibitem[\protect\citeauthoryear{Goncharov and Pavlov}{2012}]{gp12}%2
V.\,P. Goncharov and V.\,I. Pavlov,
%Blow-up instability in shallow water flows with horizontally-nonuniform density.
JETP Lett. {\bf 96}, No. 7, 427 (2012).
\url{http://dx.doi.org/10.1134/S0021364012190095}

\bibitem[\protect\citeauthoryear{Goncharov and Pavlov}{2013a}]{gp13a}%3
V.\,P. Goncharov and V.\,I. Pavlov,
%Simple model of the Rayleigh-Taylor instability, collapse, and structural elements.
Phys. Rev. E {\bf 88}, 023002 (2013).
\url{http://dx.doi.org/10.1103/PhysRevE.88.023002}

\bibitem[\protect\citeauthoryear{Goncharov and Pavlov}{2013b}]{gp13b}%4
V.\,P. Goncharov and V.\,I. Pavlov,
%Structural elements of collapses in shallow water flows.
JETP, {\bf 117}, No. 4, 754 (2013).
\url{http://dx.doi.org/10.1134/S1063776113100014}

\bibitem[\protect\citeauthoryear{Zakharov and Kuznetsov}{2011}]{zk11}%5
V.\,E. Zakharov and E.\,A. Kuznetsov,
%Solitons and collapses: two evolution scenarios of nonlinear wave systems
Physics-Uspekhi, \textbf{55}, 535 (2011)
\url{http://dx.doi.org/10.3367/UFNe.0182.201206a.0569}

\bibitem[\protect\citeauthoryear{Goncharov and Pavlov}{2006}]{gp06}%6
V.\,P. Goncharov and V.\,I. Pavlov,
%Dominant structures in jet streams
JETP Lett. {\bf 84}, 384 (2006).
\url{http://dx.doi.org/10.1134/S0021364006190064}

\bibitem[\protect\citeauthoryear{Goncharov and Pavlov}{2007}]{gp07}%7
V.\,P. Goncharov and V.\,I. Pavlov,
%Model of compactons on jet streams and their collapse
Phys. Rev. E, {\bf 76}, 066314 (2007).
\url{http://dx.doi.org/10.1103/PhysRevE.76.066314}

\bibitem[\protect\citeauthoryear{Goncharov}{2012}]{gon09}%8
V.\,P. Goncharov,
%Instability of potential jet streams
JETP Lett. {\bf 89}, 393 (2009).
\url{http://dx.doi.org/10.1134/S0021364009080049}

\bibitem[\protect\citeauthoryear{Goncharov and Pavlov}{2010}]{gp10}%9
V.\,P. Goncharov and V.\,I. Pavlov,
%Instability of gravity flows on a slope
JETP, {\bf 111}, 124 (2010).
\url{http://dx.doi.org/10.1134/S1063776110070125}

\bibitem[\protect\citeauthoryear{Goncharov}{2011}]{gon11}%10
V.\,P. Goncharov,
%Structural elements and collapse regimes in 3d flows on a slope
JETP {\bf 113}, 714 (2011).
\url{http://dx.doi.org/10.1134/S1063776111090044}

\bibitem[\protect\citeauthoryear{Goncharov and Pavlov}{2008}]{gp08}%11
V.\,P. Goncharov and V.\,I. Pavlov,
{\sl Hamiltonian Vortex and Wave Dynamics} (Geos, Moscow, 2008) (in Russian).
\url{http://dx.doi.org/10.13140/2.1.4234.3362}

\bibitem[\protect\citeauthoryear{Eggers and Fontelos}{2009}]{ef09}%12
J. Eggers and M. A. Fontelos,
%The role of self-similarity in singularities of partial differential equations.
Nonlinearity {\bf 22}, R1 (2009).
\url{http://dx.doi.org/10.1088/0951-7715/22/1/R01}

\bibitem[\protect\citeauthoryear{Kuznetsov}{2004}]{kuz04}%13
E.\,A. Kuznetsov,
%Turbulence spectra generated by singularities.
JETP Lett. {bf 80}, 83 (2004).
\url{http://dx.doi.org/10.1134/1.1804214}

\bibitem[\protect\citeauthoryear{Kuznetsov at al}{2007}]{knnr07}%14
E.\,A. Kuznetsov, V. Naulin, A.\,H. Nielsen, and J.\,J. Rasmussen,
%Effects of sharp vorticity gradients in two-dimensional hydrodynamic turbulence.
Phys. Fluids {bf 19}, 105110 (2007).
\url{http://dx.doi.org/10.1063/1.2793150}

\bibitem[\protect\citeauthoryear{Goncharov and Pavlov}{2014}]{gp14}%15
V.\,P. Goncharov and V.\,I. Pavlov,
%Whether the Sun�'s supergranulation possesses a scaling?
JETP Lett. {\bf 90}, 317 (2014).
\url{http://dx.doi.org/10.1134/S002136401406006X}

\bibitem[\protect\citeauthoryear{Goncharov and Pavlov}{2015}]{gp15a}%16
V.\,P. Goncharov and V.\,I. Pavlov,
%Effect of nonhydrostaticity on the final stage of instability in shallow water with a horizontally nonuniform density
JETP Lett. {\bf 101}, 438 (2015).
\url{http://dx.doi.org/10.1134/S0021364015070097}

\bibitem[\protect\citeauthoryear{Goncharov and Pavlov}{2015}]{gp15b}%17
V.\,P. Goncharov and V.\,I. Pavlov,
%Algebraic instability in shallow water flows with horizontally nonuniform density.
Phys. Rev. E, {\bf 91}, 043004 (2015).
\url{http://dx.doi.org/10.1103/PhysRevE.91.043004}

\bibitem[\protect\citeauthoryear{Peregrine}{1967}]{per67}%18
D.\,H. Peregrine,
%Long waves on a beach.
J. FIuid Mech. {\bf 27}, 815 (1967).
\url{http://dx.doi.org/10.1017/S0022112067002605}

\bibitem[\protect\citeauthoryear{Green and Naghdi}{1976}]{gn76}%19
A.\,E. Green and P.\,M. Naghdi,
%A derivation of equations for wave propagation in water of variable depth.
J. Fluid Mech. {\bf 78}, 237 (1976).
\url{http://dx.doi.org/10.1017/S0022112076002425}

\bibitem[\protect\citeauthoryear{Wu}{1981}]{wu81}%20
T.\,Y. Wu,
%Long waves in ocean and coastal waters.
J. Eng. Mech. Div., Proc. ASCE {\bf 107}, 501 (1981).
%Journal of the Engineering Mechanics Division, Vol. 107, No. 3, May/June 1981, pp. 501-522
\url{http://resolver.caltech.edu/CaltechAUTHORS:WUTjem81b}

\bibitem[\protect\citeauthoryear{Camassa and Holm}{1992}]{ch92}%21
R. Camassa and D.\,D. Holm,
%Dispersive barotropic equations for stratified mesoscale ocean dynamics.
Physica D {\bf 60}, 1 (1992).
\url{http://dx.doi.org/10.1016/0167-2789(92)90223-A}

\bibitem[\protect\citeauthoryear{Camassa, Holm and Hyman}{1994}]{chh94}%22
R. Camassa, D.\,D. Holm and J.\,M. Hyman,
%A new integrable shallow water equation.
Adv. Appl. Mech. {\bf 31}, 1 (1994).
\url{http://dx.doi.org/10.1016/S0065-2156(08)70254-0}

\bibitem[\protect\citeauthoryear{Ripa}{1993}]{rip93}%23
P. Ripa, %17
%Conservation laws for primitive equations models with inhomogeneous layers.
Geophys. Astrophys. Fluid Dyn. {\bf 70}, 85 (1993).
\url{http://dx.doi.org/10.1080/03091929308203588}

\bibitem[\protect\citeauthoryear{Bowman}{1987}]{b87}%24
Bowman, S.,
%A note on Hamiltonian structure.
Math. Proc. Camb. Phil. Soc. {\bf102}, 173 (1987).
\url{http://dx.doi.org/10.1007/S0305004100067165}

\bibitem[\protect\citeauthoryear{Vlasov at al}{1971}]{vpt71}%25
Vlasov, S. N., Petritshchev, V. A., and Talanov, V. I.,
%Averaged description of wave beams in linear and nonlinear media (the method of moments)
Radiophys. Quantum Electron. \textbf{14}, 1062 (1971).
\url{http://dx.doi.org/10.1007/BF01029467}

\end{thebibliography}
\end{document}